\documentstyle[psfig]{article}
\begin{document}
\huge
\begin{center}
A further note on a formal relationship between the arithmetic of homaloidal 
nets and the dimensions of transfinite space-time
\end{center}
\vspace*{.0cm}
\Large
\begin{center}
Metod Saniga

\vspace*{.3cm}
\small
{\it Astronomical Institute, Slovak Academy of Sciences, 
SK-059 60 Tatransk\' a Lomnica, Slovak Republic}
\end{center}

\vspace*{-.4cm}
\noindent
\hrulefill

\vspace*{.2cm}
\small
\noindent
{\bf Abstract}

A  sequence  of  integers  generated  by  the  number  of  conjugated pairs of 
homaloidal  nets  of  plane algebraic curves of even order is found to provide 
an {\it exact} integer-valued match for El Naschie's primordial set of fractal 
dimensions characterizing transfinite heterotic string space-time.

\noindent
\hrulefill

\vspace*{.7cm}
\large
\noindent
{\bf 1. Introduction}
\vspace*{.4cm}

Recently [1], employing the algebra and arithmetic of self-conjugate homaloidal 
nets of planar algebraic curves,  we have discovered a couple of integer-valued 
progressions  that  are  found  to  mimic extraordinary well the two El Naschie 
sequences  of  fractal  dimensions  characterizing transfinite heterotic string 
space-time.  One  of  the  progressions is demonstrated to provide an algebraic 
geometrical  justification  not  only  for  all  the relevant dimensions of the 
classical  theory,  but  also for other two dimensions advocated by El Naschie
[2--5], namely  the smallest value of the inverse of non-supersymmetric quantum
gravity coupling constant, $\bar{\alpha}_{{\rm  g}}= 42+2k \simeq 42.36067977$,
and the inverse of fine structure constant, $\bar{\alpha}_{0}=2 \otimes (68+3k)
\simeq 2 \otimes 68.54101967$.  A  non-trivial  coupling  between  the  two  El 
Naschie sequences has also been revealed.  

\vspace*{.7cm}
\noindent
{\bf 2. Conjugate pairs of homaloidal nets and primordial El Naschie's 
dimensions}
\vspace*{.4cm}

In the present paper, we put forward another `homaloidal' sequence that offers
a  nice  fit  to  the  primordial  set  of fractal dimensions introduced by El 
Naschie, i.e. to the set defined by [2--5]
\begin{equation}
D_{q}=[136+6\phi^{3}\left(1-\phi^{3}\right)] \phi^{q}
\end{equation}
where $q$  is a non-negative integer and  $\phi$ is taken to be identical with 
the  golden  mean,  $\phi= 1 - \phi^{2} = \left(\sqrt{5} - 1 \right)/ 2 \simeq  
0.618033989$.  The  new sequence is borne by the numbers of conjugate pairs of 
homaloidal  nets  of a given order $n$, $\#^{{\rm pair}}_{n}$ (see [1,6,7] for 
the definitions,  symbols  and notation). From the properties of plane Cremona 
transfomations  [6--8]  it  is  obvious  that  these  numbers are given by the 
following simple formula
\begin{equation}
\#^{{\rm pair}}_{n} = \frac{\#^{{\rm tot}}_{n} - \#^{{\rm sc}}_{n}}{2} +
\#^{{\rm sc}}_{n} = \frac{\#^{{\rm tot}}_{n} + \#^{{\rm sc}}_{n}}{2}
\end{equation}
with  $\#^{{\rm tot}}_{n}$  standing for the total number of nets of order $n$
and  $\#^{{\rm sc}}_{n}$  representing  the  number  of  those  nets  that are 
self-conjugate.  As  in  the  preceding  paper  [1], our homaloidal numerology 
relies  almost  exclusively on the `data' found in an old, but very important, 
paper  by  B. Mlodziejowski [8], where the values of both $\#^{{\rm tot}}_{n}$ 
and $\#^{{\rm sc}}_{n}$ are given for the first 22 orders. Exploiting Eq. (2), 
we  have  obtained the values of $\#^{{\rm pair}}_{n}$, $n$ even, as listed in 
column  two  of  the  table  below.  The  sequence  itself is generated by the 
`augmented-by-one' quantities, 
\begin{equation}
\widehat{\#}^{{\rm pair}}_{n} \equiv \#^{{\rm pair}}_{n} + 1;
\end{equation}
these  are given in column three of the table. We see that this sequence is an 
{\it  exact}  integer-valued  fit of the El Naschie sequence (introduced in an
explicit form in the last column of the table); that is, the absolute value of 
the difference between $\widehat{\#}^{{\rm pair}}_{n}$  and  its corresponding
fractal dimension $D_{q}$ is always less than {\it one}!

\vspace*{.4cm}
\begin{tabular}{llll}
\hline \hline 
~~$n$~~~~ & $\#^{{\rm pair}}_{n}$ & $\widehat{\#}^{{\rm pair}}_{n}$ & 
$D_{q}$ (El Naschie's primordial dimensions) \\
\hline  
~~2 & 1 & 2 & $D_{q=9} = 0 + 10k \simeq 1.803\ldots $ \\
~~4 & 2 & 3 & $D_{q=8} = 4 - 6k \simeq 2.917\ldots $ \\
~~6 & 3 & 4 & $D_{q=7} = 4 + 4k \simeq 4.721\ldots $ \\
~~8 & 7 & 8 & $D_{q=6} = 8 - 2k \simeq 7.639\ldots $ \\
~~10 & 11 & 12 & $D_{q=5} = 12+2k \simeq 12.360\ldots $ \\ 
~~12 & 19 & 20 & $D_{q=4} = 20 + 0k = 20 $ \\  
~~14 & 31 & 32 & $D_{q=3} = 32+2k \simeq 32.360\ldots $ \\  
~~16 & 51 & 52 & $D_{q=2} = 52+2k \simeq 52.360\ldots $ \\  
~~18 & 84 & 85 & $D_{q=1} = 84+4k \simeq 84.720\ldots $ \\
~~20 & 137 & 138 & $D_{q=0} = 136+6k \simeq 137.082\ldots$ \\ 
~~22 & 202 & 203 &  no match ($D_{q=-1} = 220+10k 
\simeq 221.803\ldots $) \\
\hline \hline 
\end{tabular}
\vspace*{.4cm}

What strikes most about the sequence $\widehat{\#}^{{\rm pair}}_{n=2l}$, $l=1, 
2,\ldots$, when compared with  its `self-conjugate' counterpart $D^{(2)}_{l}$ 
described in [1],  are the following two facts: 1) it copies $D_{q}$ within a 
{\it greater}  interval  of values of $l$ ($1 \leq l \leq 10$, whereas in the 
case of $D^{(2)}_{l}$  it  is  for  $1 \leq l \leq 6$  only [1]); and  2) the 
correspondence {\it terminates} at $D_{q=0} \simeq 137.082\ldots$ -- the value 
that is believed to be very close  to the inverse value of the fine structure 
constant [2,9].  It  is especially the latter fact which may be regarded as a 
further justification of the  prominent  role this  constant  is  supposed to 
play in all fundamental theories of stringy  space-times  and Cantorian space 
${\cal E}^{(\infty)}$ as well [9].

\vspace*{.7cm}
\noindent
{\bf 3. Conclusion}
\vspace*{.4cm}

The theory outlined in [1] and the present paper strongly suggests that we 
could gain some important physical insights into the nature of transfinite
heterotic string space-time making use of the concepts of homaloidal nets and 
plane Cremona transformations. It is our hope that the two papers will elicit 
the interest of other physicists to explore its further possibilities so that 
we may soon know whether the extremely exciting prospects implicit in this 
rigorous arithmetic-algebraic approach are real or illusory.

\vspace*{.7cm}
\noindent
{\bf Acknowledgements}
\vspace*{.4cm}

I am extremely indebted to Mrs. Bella Shirman, a senior librarian of the
University of California at Berkeley, for her kind assistance in tracing and 
making for me a copy of Ref. 8. This work was supported in part by the NATO 
Collaborative Linkage Grant PST.CLG.976850.
\\ \\ \\

\end{document}